\documentclass[onecolumn,useAMS,usenatbib]{mn2e}

\usepackage{color}
\usepackage{amsmath}
\usepackage{graphicx}
\usepackage{amssymb}
\usepackage{psfrag}

\newcommand{\beq}{\begin{equation}}
\newcommand{\eeq}{\end{equation}}
\newcommand{\bay}{\begin{array}}
\newcommand{\eay}{\end{array}}
\newcommand{\beqy}{\begin{eqnarray}}
\newcommand{\eeqy}{\end{eqnarray}}
\newcommand{\nn}{\nonumber}
\newcommand{\rmd}{\mathrm{d}}
\newcommand{\brac}[1]{\left({#1}\right)}

\newcommand{\pd}[2]{\frac{\partial{#1}}{\partial{#2}}}
\newcommand{\td}[2]{\frac{\rmd{#1}}{\rmd{#2}}}
\newcommand{\curl}{\nabla\times}
\renewcommand{\div}{\nabla\cdot}
\newcommand{\be}{{\bf e}}
\newcommand{\br}{{\bf r}}

\newcommand{\bj}{{\bf j}}
\newcommand{\bB}{{\bf B}}
\newcommand{\pom}{\varpi}

\newcommand{\Pp}{\mathcal{P}^+}
\newcommand{\Pm}{\mathcal{P}^-}

\newcommand{\clE}{\mathcal{E}}
\newcommand{\emag}{\clE_{\textrm{mag}}}
\newcommand{\etor}{\clE_{\textrm{tor}}}
\newcommand{\bOm}{{\boldsymbol{\Omega}}}

\newcommand{\bv}{{\bf{v}}}

\newcommand{\drh}{\delta\rho}

\newcommand{\boldf}{{\bf f}}

\newcommand{\smax}{\textrm{surfmax}}
\newcommand{\gmax}{\textrm{globmax}}

\newcommand{\skl}[1]{{\color{black}{#1}}}

\title[Magnetic stability of barotropic stars]{Are there any stable magnetic
  fields in barotropic stars?}
\author[S. K. Lander and D. I. Jones]
       {S. K. Lander${}^{1,2}$\thanks{samuel.lander@uni-tuebingen.de}
         and D. I. Jones${}^1$\\ \\
${}^1$ School of Mathematics, University of Southampton, Southampton
       SO17 1BJ, U.K.\\
${}^2$ Theoretical Astrophysics, University of T\"ubingen, Auf der
Morgenstelle 10, T\"ubingen 72076, Germany}

\begin{document}

\pagerange{\pageref{firstpage}--\pageref{lastpage}} \pubyear{0000}
\maketitle

\label{firstpage}

\begin{abstract}
We construct barotropic stellar equilibria, containing magnetic
fields with both poloidal and toroidal field components. We extend earlier results by
exploring the effect of different magnetic field and current
distributions. Our results suggest that the boundary treatment plays a
major role in whether the poloidal or toroidal field component is
globally dominant. Using time evolutions we provide the first
stability test for mixed poloidal-toroidal fields in barotropic stars,
finding that all these fields suffer instabilities due to one of the field
components: these are localised around the pole for toroidal-dominated
equilibria and in the closed-field line region for poloidal-dominated
equilibria. Rotation provides only partial stabilisation. There appears to be very limited
scope for the existence of stable magnetic fields in barotropic
stars. We discuss what additional physics from real stars may
allow for stable fields.
\end{abstract}


\section{Introduction}

At least three different classes of star appear to harbour strong and broadly
similar magnetic fields: Ap/Bp stars on the main sequence, magnetic
white dwarfs and neutron stars.  All three classes exhibit long-lived
fields, with large-scale simple structure. Surface fields are up to
the order of $10^4$ G for 
Ap/Bp stars, $10^9$ G for magnetic white dwarfs and $10^{15}$ G for
neutron stars. Although these values are very different, the resultant
total magnetic flux is similar for the three classes, due to their
correspondingly different radii. Given their high strengths, magnetic fields are
expected to play important roles in the 
evolution and dynamics of these stars
\citep{don_land,wick_ferr,harding_lai}. In addition, magnetic fields
distort a star \citep{chand_fermi}; this distortion may (in a rotating
neutron star) lead to appreciable emission of gravitational waves \citep{bon_gour}.

\skl{Observations only give us direct information about the
  exterior field of a star, although it may be possible to infer
  details of the interior field of neutron stars from magnetar flares
  and oscillations (at present) and through gravitational-wave
  emission (in the future). Having no direct probe of the interior} is a problem, since the
details of the field geometry can greatly affect the ratio of
internal-to-external field strength; the observed field may provide a
poor estimate of the global value. For example, if a star's magnetic field is
confined to an outer region (like the crust of a neutron star), its
volume-averaged value could be lower than expected from the observed
surface value; if the field is predominantly buried within the star,
its strength may be greater than expected. Innumerable studies have
been motivated by the attempt to understand the kinds of magnetic
field that may exist in stars (see \citet{mestel-book} for a review).

One approach to the modelling of stellar magnetic fields is to
construct equilibrium configurations of a fluid star, and then to
show that they are dynamically stable (and hence credible models for
the long-lived magnetic fields present in many real stars). In practice, the latter
step is very difficult --- it requires confirmation that
\emph{every} possible perturbation about a stellar equilibrium is
stable. For the former step, constructing hydromagnetic equilibria,
virtually all studies make the simplifying assumption of a barotropic
(i.e. unstratified) stellar model: the pressure is taken as a function of density
alone. This has been
criticised as unrealistic --- temperature/entropy profiles provide stratification
in main-sequence stars \citep{mestel_strat} and white dwarfs
\citep{reis_strat}, whilst the varying proportions of charged
particles (composition gradients) result in neutron-star
stratification \citep{reis_gold}. Stratification provides an additional degree of stabilisation, and
numerical simulations for main-sequence stars with entropy 
gradients have provided strong evidence for the existence of stable
magnetic equilibria in this case \citep{braith_nord,braith_torpol}. It
is not clear if this result is applicable to neutron stars, however,
where the origin of the stratification is different.

Despite its limitations, a barotropic equation of state is a sensible
first approximation to stellar matter, and one which is
astrophysically relevant if it allows for stable magnetic
equilibria. It is still unknown whether such stable configurations
exist, but there seem two reasons to expect them to. The first is
connected with the geometry of a mixed poloidal-toroidal field. A
poloidal field is known to be unstable in the region where its field
lines close within the star, whilst a toroidal field suffers localised
instabilities around the pole; in a mixed field, each field component
`fills in' the region where the other field component is
unstable. This changes the magnetic field geometry in these
unstable regions, potentially suppressing the instabilities
\citep{wright,taylermix}. The second reason to expect stable
equilibria is that the `twisted-torus' fields found for barotropic
equilibria \citep{LJ09,ciolfi_eqm} are qualitatively similar to the
final state of the nonlinear stratified-star simulations mentioned
above \citep{braith_nord}.

This paper aims to explore the possibility of stable magnetic fields
existing in barotropic stars, by testing the stability of a number of
equilibrium models. We begin, in section 2, by exploring the range of
magnetic equilibria possible 
in barotropic stars, reviewing the key equations of the problem and
examining the various classes of solution. Broadly, the resulting equilibria
differ in how their magnetic field and current are distributed. In section 3, we discuss
properties of perturbations in a magnetic star: the governing
equations, how a mixed poloidal-toroidal field breaks the equatorial
symmetry of the perturbations, and potential instabilities. Next, we use the
mixed-field equilibria of section 2 as background configurations for
time evolutions of the linearised perturbation equations (section
4). We study perturbations known to result in instabilities for purely
poloidal/toroidal fields, to determine whether or not these
perturbations are still unstable in mixed-field stars. Following this,
we discuss some important aspects of stellar physics not accounted for in
barotropic models, and whether these may remove magnetic instabilities
(section 5). Finally, in section 6 we summarise our results.

\section{Mixed-field barotropic equilibria}

The first step towards studying the stability of magnetised stars is
to construct suitable background equilibrium models, which we describe in this
section.  Although full details may be found in \citet{LJ09}, we pause
here to focus on the key equations of the problem, and any
restrictions imposed by our assumptions. This is
important, since our aim is to produce the widest possible range 
of equilibrium models to test for stability.

We model a star as a perfectly-conducting barotropic fluid body, in Newtonian gravity and
axisymmetry. The star has a mixed poloidal-toroidal magnetic field $\bB$,
and our scheme also allows for rigid rotation (about the
magnetic symmetry axis) with angular velocity $\Omega$. We work in
cylindrical polar coordinates 
$(\varpi,\phi,z)$, with the $z$-axis being aligned with the symmetry
axis of the star. An equilibrium is described by the stationary
form of the MHD Euler equation:
\beq \label{eqm_Euler}
\frac{\nabla P}{\rho}+\nabla\Phi
  -\nabla\brac{\frac{\varpi^2\Omega^2}{2}}
  -\frac{\bj\times\bB}{\rho} = 0,
\eeq
where $P$ is fluid pressure, $\rho$ mass density, $\Phi$ gravitational
potential and $\bj$ the electric current. This equation needs to be
solved together with Amp\`ere's law
\beq \label{ampere}
4\pi\bj=\curl\bB,
\eeq
Poisson's equation
\beq
\nabla^2\Phi=4\pi G\rho,
\eeq
where $G$ is the gravitational constant; and the solenoidal constraint on the magnetic field
\beq
\div\bB=0.
\eeq
The system is closed with an equation of state. We choose a polytrope
with index $N=1$, a typical approximation for neutron star
matter:
\beq
P=P(\rho)=k\rho^{1+1/N}=k\rho^2.
\eeq
Note that a more suitable value for white dwarfs would be $N=1.5$,
whilst main sequence stars tend to be modelled as $N=3$
polytropes. We are able to generate magnetic equilibria for a variety
of values of $N$ \citep{LJ09}, but the resulting field configurations
are all qualitatively similar. Since magnetic-field instabilities are
related to the geometry of the field (see section
\ref{instabs_theory}), we have some expectation that our results
for $N=1$ should be representative of the stability of barotropic
main-sequence stars and white dwarfs, not just neutron stars.

Some algebra is needed to recast the equilibrium equations above into a convenient
form for solution. First let us take the curl of the Euler equation
\eqref{eqm_Euler}, which --- in the case of a \emph{barotropic} star --- yields
\beq
\frac{\bj\times\bB}{\rho} = \nabla M,
\eeq
for some scalar function $M$.
The solenoidal constraint on the magnetic field together with the
assumption of axisymmetry allows us 
to express the poloidal field component $\bB_{pol}$ in terms of a
streamfunction $u$:
\beq
\bB_{pol}=\frac{1}{\varpi}\nabla u\times\be_\phi.
\eeq
Furthermore, the toroidal component is related to the streamfunction
through some function $f$:
\beq
B_\phi = \frac{1}{\varpi}f(u)
\eeq
and it may also be shown that $M=M(u)$. So, the magnetic field and the
force it exerts on the fluid are both related to the single scalar
function $u$. Equivalently, it is possible to work with the
$\phi$-component of the magnetic vector potential ${\bf A}$ instead, 
since $u=\varpi A_\phi$. With various algebraic tricks, one may derive a relation between these
functions $u,M$ and $f$, known as the Grad-Shafranov equation \citep{grad_rubin,shafranov}:
\beq
-4\pi\rho\varpi^2\td{M}{u}
 = f\td{f}{u} + \brac{\pd{^2}{\varpi^2}-\frac{1}{\varpi}\pd{ }{\varpi}+\pd{^2}{z^2}}u.
\eeq
Combining the Grad-Shafranov equation with Amp\`ere's law
\eqref{ampere} in
axisymmetry yields a useful relation between the magnetic functions,
the current and the field:
\beq \label{current}
\bj = \frac{1}{4\pi}\td{f}{u}\bB
           + \rho\varpi\td{M}{u}\be_\phi.
\eeq
By rewriting $\bj$ and $\bB$ in terms of the streamfunction
\citep{tomi_eri} we arrive at a
version of Poisson's equation for the magnetic field:
\beq \label{deriv_uint}
\nabla^2\brac{\frac{1}{\varpi}u\sin\phi}
 = -\brac{\frac{f}{\varpi}\td{f}{u}+4\pi\varpi\rho \td{M}{u}}\sin\phi.
\eeq
The above result is the last that one can obtain without loss of
generality. For numerical solution we now need an integral form of equation
\eqref{deriv_uint}, and in choosing a Green's function to  
produce this integral we also implicitly choose a boundary
condition for the exterior field. All of the equilibria we
generate use the integral equation
\beq \label{int_uint}
\frac{1}{\varpi}u\sin\phi
 = \frac{1}{4\pi}\int
         \frac{f'(\tilde{u})f(\tilde{u})/\tilde\varpi
           +4\pi\tilde\varpi\tilde\rho M'(\tilde{u})}{|\br-\tilde\br|}
                \sin\tilde\phi\ \rmd\tilde\br,
\eeq
for which the magnetic-field magnitude falls off as a dipole, $B\to
1/r^3$ as $r\to\infty$. If one wished instead to solve for fields
confined within the star (for example), \skl{a modified Green's
  function would be necessary.}

One must also specify the functional forms of $M(u)$ and $f(u)$, which
are chosen on physical grounds.  In particular, we see from equation
\eqref{current} that the derivatives $M'(u)$ and $f'(u)$ are
related to the distribution of electric current. Both $M$ and $f$ can
be chosen to allow for smooth or discontinuous current distributions
inside the star, whilst most choices for $f$ result in currents
outside the star. \skl{As discussed later, the limited choices of $f(u)$ which
avoid exterior currents all produce qualitatively similar equilibria.}

The integral equations for $\Phi$ and $u$ are solved together with the
Euler equation using a numerical scheme which iteratively finds equilibrium
configurations --- for details see \citet{LJ09} or
\citet{tomi_eri}. \skl{For numerical solution, it is convenient to use
  dimensionless variables, which we denote by a hat
  (e.g. $\hat\Omega$). We produce these by dividing 
  each physical quantity by the requisite combination of powers of
  $G$, maximum density $\rho_{max}$ and equatorial radius
  $r_{eq}$. The results of this section are presented in dimensionless
  form, to avoid specialising to a specific star. In addition, the important
  features of the equilibria --- the distribution and relative strength of the
  two field components --- are clear without redimensionalising.} We
now consider different choices of the two magnetic functions $f(u)$
and $M(u)$ and the resultant equilibria.

\subsection{Exterior poloidal field, no surface currents; `type 1'}
\label{tt_eqa}

Our first choice of surface treatment is that employed in \citet{LJ09}.
Specifically, the toroidal component is confined within the star and
goes to zero smoothly at its surface, whilst the internal poloidal
field matches smoothly to an external component, which falls off as
$1/r^3$. We feel this is the most natural boundary condition for a
fluid star in vacuum, although it ignores the different physics
present in the outer regions of stars --- for example the crust,
ocean and magnetosphere of a neutron star. One surprising, perhaps
unsatisfactory, aspect of these solutions is that we find the toroidal
component is always weak with respect to the poloidal one in a global sense; its
contribution $\etor$ to the magnetic energy $\emag$ is never more than
a few per cent of the total. Locally the two field components can be
more similar, however, with the maximum values of poloidal and toroidal
components being of comparable magnitude.

To produce the desired surface behaviour of the toroidal field
component, we need to make a suitable choice of magnetic function
$f(u)$. \skl{We take}
\beq \label{fchoice_case1}
\skl{f(u)} = \begin{cases}
                      a(u-u_\smax)^\skl{1.1}  & u>     u_\smax\\
                      0                                     & u\leq u_\smax,
                   \end{cases}
\eeq
where $u_\smax$ is the maximum surface value attained by $u$, as
discussed by various authors
\citep{tomi_eri,LJ09,ciolfi_eqm2,lyut}, \skl{and $a$ is a constant
  coefficient related to the relative strength of toroidal and
  poloidal components ($a=0$ produces a purely poloidal field). Experimenting with different
exponents of $(u-u_\smax)$, we have found that lower values allow for slightly
stronger toroidal components. At the same time, we wish to avoid
producing a step in $f'(u)$, so we need the exponent to be greater than
unity; hence we chose it as $1.1$. In other cases the equilibria are
still qualitatively similar.} The
functional form \eqref{fchoice_case1} has the effect of enclosing the
toroidal field component within the closed-field line region of the
star, producing equilibria where the toroidal field only occupies a
small volume of the star. Nonetheless, it is the largest volume that
does not give rise to an exterior current.

\skl{Within an iterative scheme,} we consider this choice slightly inconsistent with the requirement
that $f$ be a function of $u$ alone: $u_\smax$ is a constant, but one
found by evaluating $u$ at $\rho=0$. The value of $u_\smax$ varies between
iterative steps, and hence the scheme involves an implicit
density dependence. Instead, it is possible to produce entirely
consistent equilibria using the streamfunction's global maximum 
$u_\gmax$ (which is attained within the star): one can fit the $f$
function to contours of different fractions of $u_\gmax$ and pick the
solution with the largest volume of toroidal component which does not
extend outside the star. The resulting
equilibria appear identical to those using $u_\smax$, however, 
suggesting that \eqref{fchoice_case1} is in fact an acceptable choice.

\begin{figure}
\begin{center}
\begin{minipage}[c]{\linewidth}
\psfrag{1a}{type 1a}
\psfrag{1b}{type 1b}
\psfrag{1c}{type 1c}
\psfrag{Br}{\small{$B_r^{\rm pole}$}}
\psfrag{Bth}{\small{$B_\theta^{\rm eq}$}}
\psfrag{Bph}{\small{$B_\phi^{\rm eq}$}}
\psfrag{magB}{\large{$\hat{B}$}}
\psfrag{r/R}{\large{$r/R_*$}}
\includegraphics[width=\linewidth]{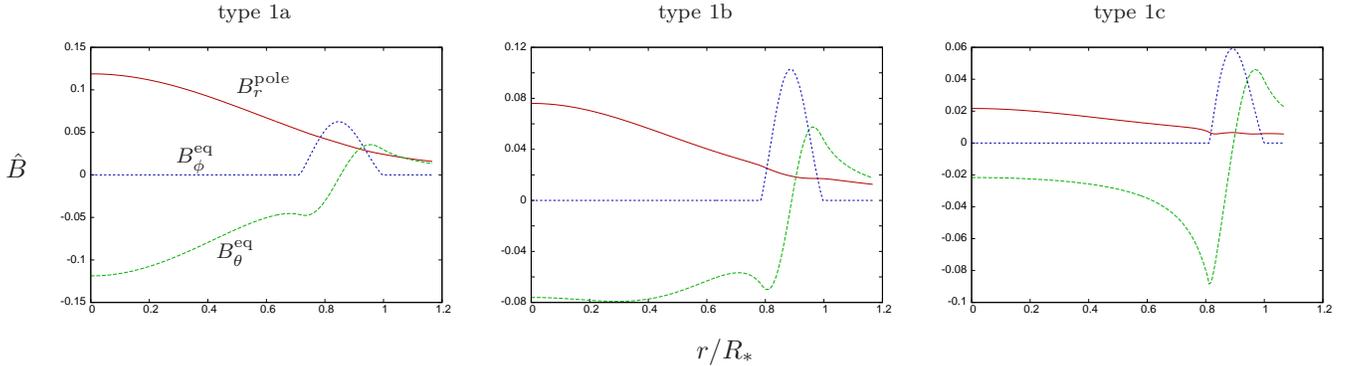}
\end{minipage}
\caption{\label{type1}
               Three different type 1 or `twisted-torus' equilibria,
               all with a relatively large toroidal component:
               $\etor/\emag=3\%,4.5\%,4.5\%$ for types 1a, 1b and 1c respectively.
               The variation in magnitude of the three field components is shown as
               a function of dimensionless stellar radius $r/R_*$,
               where $R_*$ is the surface of the star. $B_\theta$
               and $B_\phi$ are evaluated along the equatorial axis,
               whilst $B_r$ (zero along this axis) is evaluated along the
               polar axis. The neutral line of the poloidal component
               is located where the $B_\theta$ line crosses the
               $x$-axis, and the toroidal component is centred around
               this line. These three equilibria represent the range
               of solutions we have found which have an exterior
               field but no surface/exterior current; nonetheless, all
               are qualitatively similar.}
\end{center}
\end{figure}

All of our type-1 equilibria use the above choice for $f$, but
differ in the choice of the other magnetic function $M$. In \citet{LJ09} we
found that only $M(u)=\kappa u$ (where $\kappa$ is a constant) produced magnetised
equilibria, with other cases resulting in our numerical scheme iterating to unmagnetised
solutions; here we refer to the $M=\kappa u$ equilibria as type
1a. \skl{Physically, $\kappa$ gives the relative strength of the Lorentz
force compared with gravity and is thus related to the magnetic field
strength; setting $\kappa=0$ produces an unmagnetised star.}

More recently, we have found a way of using other functional forms of
$M$ without the numerical scheme iterating to a zero-field
solution. By including a $1/u_\gmax$ factor, the field at each
iterative step is `normalised' to the same magnitude as the previous
step and not driven to zero. These different functional forms of $M$
produce qualitatively similar equilibria, so we take one
representative example --- $M=\kappa u^2/u_\gmax$ --- and refer to it as type 1b.

A different generalisation of the original type-1a equilibria is to
relax the requirement that the current \skl{density} be continuous inside the
star. Allowing for steps in the current means we can choose forms of
$M$ and $f$ with discontinuous $u$-derivatives. We find that a step in
$f'(u)$ produces little qualitative difference in equilibria from
types 1a or 1b, so we do not consider these further. More significant
differences emerge by choosing a step in $M'(u)$ --- as an example of
these we choose the following:
\beq
M'(u) = \begin{cases}
              \kappa  & u>     u_\smax\\
              0           & u\leq u_\smax.
           \end{cases}
\eeq
Note that this corresponds to a step in the interior current, but not
to a surface current; see equation \eqref{current}. Since we wish to
avoid $\delta$-distribution behaviour of $M'(u)$ (and hence the
Lorentz force) in this class of solutions, however, we must ensure that
$M$ itself is smooth --- and so we take
\beq
M(u) = \begin{cases}
              \kappa(u-u_\smax)  & u>     u_\smax\\
              0                               & u\leq u_\smax.
           \end{cases}
\eeq
\skl{This choice confines the Lorentz force within the closed-field
  line region, with a force-free field existing in the rest of the
  stellar interior as well as the exterior.} The other magnetic function $f$
is chosen in the same way as for type 1a. This final choice we refer to as
type 1c.

The three type-1 field configurations are compared in figure \ref{type1},
where we plot the variation in magnitude of $B_r,B_\theta$ and
$B_\phi$ with stellar radius.  The maximum value attained by the
toroidal field is greater in type 1b than type 1a, but it is still
confined to the same small region.  The profiles of types 1a and 1c differ
mainly in the relatively weak radial field in type 1c.  Overall,
however, the three field configurations appear very similar.

\subsection{Confined fields, no exterior poloidal component; `type 2'}
\label{cf_eqa}

Next, we consider a very different class of equilibrium from that of the previous
subsection: one where the magnetic field is confined to the stellar
interior.  We will refer to such confined-field equilibria as `type 2'.
As described earlier, the boundary condition on the magnetic field is
imposed by the choice of Green's function used to solve equation
\eqref{deriv_uint}; to generate equilibria with confined fields we would
need to modify this function. Since the total magnetic field
vanishes at the stellar surface in this case, there would be no separate need
to ensure that the toroidal field vanishes outside the star.

\begin{figure}
\begin{center}
\begin{minipage}[c]{0.35\linewidth}
\psfrag{Br}{\small{$B_r^{\rm pole}$}}
\psfrag{Bth}{\small{$B_\theta^{\rm eq}$}}
\psfrag{Bph}{\small{$B_\phi^{\rm eq}$}}
\psfrag{magB}{\large{$\hat{B}$}}
\psfrag{r/R}{\large{$r/R_*$}}
\includegraphics[width=\linewidth]{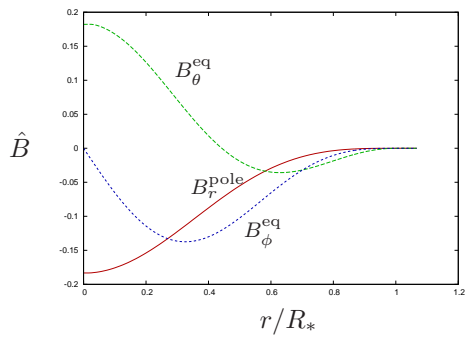}
\end{minipage}
\caption{\label{type2}
               A type-2 equilibrium, where all field components are
               confined within the star, not just $B_\phi$. Plotted
               here is the lowest-eigenvalue solution of
               \citet{haskell}, corresponding to no nodes in the
               toroidal field. 65\% of the magnetic energy is
               contained in the toroidal component --- far higher than
               for type-1 equilibria.}
\end{center}
\end{figure}

The major disadvantage in assuming a confined field is the more
limited astrophysical relevance: in particular, many stars have strong
exterior magnetic fields, thought to be comparable in strength with
the internal value. One possible application of this treatment could
be to model accreting neutron stars or white dwarfs, where the surface field is weak but a
strong internal field could still exist, `buried' under the accreted
matter.

There are interesting differences between the properties of magnetic
equilibria in the confined and non-confined cases. The
non-confined models described above (type 1) are `poloidal-dominated' ---
the poloidal field component contains most of the star's magnetic
energy --- whilst confined-field solutions seem to be
toroidal-dominated by the same measure \citep{ioka,haskell,duez}. Related to
this, type-1 magnetic fields induce oblate distortions in the star's
density distribution, whereas type-2 equilibria have prolate
ellipticities. Finally, since the magnetic fields in the two cases are
qualitatively different, it is natural to expect them to have
different stability properties.

Generating confined-field equilibria would not be a straightforward
extension to our work, nor is there strong motivation to do so from
observations. On the other hand, we would like to compare the two
contrasting field geometries of type 1 and 2 equilibria.  To this
end, we turn to the perturbative semi-analytic models in section B2 of
\citet{haskell} for
confined mixed-field stars. \skl{Their magnetic field (in spherical
  polar coordinates) is of the form
\beq
\bB = \frac{2A\cos\theta}{r^2}\be_r
           - \frac{A\sin\theta}{r}\be_\theta
           + \frac{\pi\lambda A\sin\theta}{r R_*}\be_\phi,
\eeq
with $A$ a radial function given by
\beq
A(r) = \frac{B_k R_*^2}{(\lambda^2-1)^2 y}
           \left[  2\pi\frac{\lambda y\cos(\lambda y)-\sin(\lambda y)}
                                    {\pi\lambda\cos(\pi\lambda)-\sin(\pi\lambda)}
                      +\left[(1-\lambda^2)y^2-2\right]\sin y + 2y\cos y
                    \right],
\eeq
where $y=\pi r/R_*$ and $B_k$ is a constant governing the field
strength. By demanding that the exterior field vanish and the 
interior field be finite and continuous, this becomes an eigenvalue problem
to solve for $\lambda$, with only a discrete set of solutions being
admissible. Higher eigenvalues correspond to a stronger toroidal
component with an increasing number of nodes. \citet{haskell} found
that the lowest-eigenvalue solution for the magnetic field was
$\lambda=2.362$, corresponding to a nodeless toroidal component.} We
adopt this field configuration together with a spherical density
distribution as our `type 2' equilibrium, neglecting the (small)
distorting effect of the magnetic field. Unlike our type-1 equilibria,
this is not fully self-consistent, but the resulting models should be
accurate enough for a qualitative idea of stability. 

We plot this `type 2' magnetic field in figure
\ref{type2}.  Even for this lowest-eigenvalue solution, the
toroidal-component magnetic energy is 65\% of the total\footnote{Table
  1 from \citet{haskell} indicates that $\etor/\emag=90\%$, due to a
  normalisation error: their averaged poloidal field is $\sim
  2.2\times 10^{12}$ G, not $1\times10^{12}$ G as reported.}.
Clearly, the toroidal field is far more globally significant than for
type-1 equilibria, extending throughout the stellar interior.

\subsection{Exterior poloidal component, surface currents; `type 3'}
\label{sc_eqa}

The types of magnetic field outlined in the last subsections
represent two extremes of mixed-field equilibria with a confined toroidal
component; in type 1 the poloidal component is non-zero across the
surface and into the exterior, whilst in type 2 it vanishes at the
surface. For both types, the toroidal component goes to zero smoothly
at the surface.  Type 1 equilibria 
are always dominated by the poloidal component; type 2 by the toroidal
component. Since we wish to test the dependence of stability on the
relative strength of the two field components, we would like a way of
introducing a `sliding scale' between types 1 and 2, to produce
equilibria with more equal proportions of each field component.

As a way to produce mixed-field equilibria with stronger toroidal
components as well as an exterior field, we consider equilibria with surface currents in
this subsection. We stress that we are not asserting that surface currents
themselves are necessarily significant in real stars --- they simply provide a
mathematically convenient way of producing a surface boundary
condition which is different from that of the type 1 or 2 equilibria
discussed above. \skl{There is some motivation for employing a
  boundary condition `in between' that of types 1 and 2, however:
  in the outer region of a star the mass density becomes very low and
  the resistivity increases, so deviations may be expected from an
  ideal-MHD treatment. Furthermore, modelling a neutron star as a
  fluid body in vacuum neglects the effect of its crust, ocean
  and magnetosphere. A first, crude attempt to account for these
  differences could be to modify the boundary condition at the stellar
  surface, which may mathematically (though not physically) resemble a
  surface current.}

\begin{figure}
\begin{center}
\begin{minipage}[c]{0.7\linewidth}
\psfrag{type 3a}{type 3a}
\psfrag{type 3b}{type 3b}
\psfrag{Br}{\small{$B_r^{\rm pole}$}}
\psfrag{Bth}{\small{$B_\theta^{\rm eq}$}}
\psfrag{Bph}{\small{$B_\phi^{\rm eq}$}}
\psfrag{magB}{\large{$\hat{B}$}}
\psfrag{r/R}{\large{$r/R_*$}}
\includegraphics[width=\linewidth]{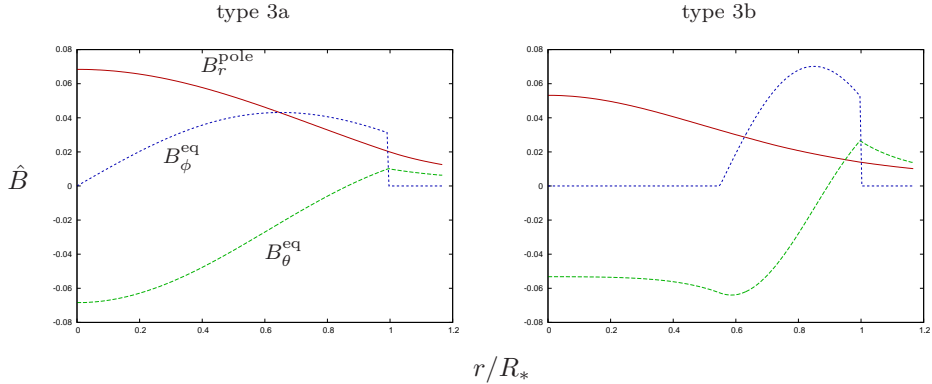}
\end{minipage}
\caption{\label{type3}
               Equilibria with the toroidal field component matched to
               a surface current rather than going smoothly to zero
               there. For the type 3a model, $\etor/\emag=50\%$; for type 3b it is 30\%. The
               right-hand plot is an attempt to produce an equilibrium
               with a surface current that also looks roughly like a
               twisted-torus structure.}
\end{center}
\end{figure}

For these equilibria we choose a form of $f$ such that $B_\phi$ is
non-zero up to the stellar surface. An exterior toroidal field requires
an exterior current; since we still want to avoid this we appeal to
some (poloidal) surface current to match the non-zero interior
toroidal field to a zero exterior one. This surface current is not
separately modelled; its form is whatever required to produce a
consistent matching for the toroidal field component. A similar
approach to the construction of magnetic equilibria was taken by
\citet{colaiuda}. \skl{We caution the reader that this surface current
  will, in general, induce an azimuthal component of the Lorentz force
  at the surface; in reality, this would need to be balanced by other physics beyond our
  fluid star model.} In practice, for this class of equilibria only the form of
the magnetic function $f$ is changed. One simple choice is:
\beq
f'(u) = \begin{cases}
              a  & {\rm interior}\\
              0  & {\rm exterior}
           \end{cases}
\eeq
\beq
f(u) = \begin{cases}
              au  & {\rm interior}\\
              0    & {\rm exterior},
           \end{cases}
\eeq
which we refer to as type 3a. Note that since $f(u)=\pom
B_\phi$, this corresponds to a jump in the toroidal field component at
the stellar surface.

The toroidal field is now distributed throughout the stellar
interior. As a half-way choice between this type of equilibrium and
the usual twisted-torus structures of type 1 (where the toroidal
field only exists in some small torus near the surface), we consider one final
class of equilibrium --- type 3b --- designed to look like a twisted-torus
configuration but with a larger torus than in type 1. Again, we use a surface current to
achieve this:
\beq
f'(u) = \begin{cases}
              a                         & \textrm{interior \underline{and} } u>0.5u_\smax\\
              0                         & \textrm{otherwise}
           \end{cases}
\eeq
\beq
f(u) = \begin{cases}
              a(u-0.5u_\smax)    & \textrm{interior \underline{and} } u>0.5u_\smax\\
              0                             & \textrm{otherwise}.
           \end{cases}
\eeq

The relative magnitudes of the $r,\theta$ and $\phi$ components of the
magnetic field in these surface-current equilibria are plotted in
figure \ref{type3}. Type-3a equilibria look quite similar to the
confined-field solution in figure \ref{type2}, differing mostly near
the surface, since the surface-current equilibria have exterior
fields. The type-3b equilibrium was constructed to resemble a
twisted-torus (i.e. type-1) equilibrium, but with a larger toroidal
component; cf. figure \ref{type1}.

\begin{figure}
\begin{center}
\begin{minipage}[c]{0.6\linewidth}
\psfrag{type 1a}{type 1a}
\psfrag{type 2}{type 2}
\psfrag{type 3a}{type 3a}
\psfrag{type 3b}{type 3b}
\includegraphics[width=\linewidth]{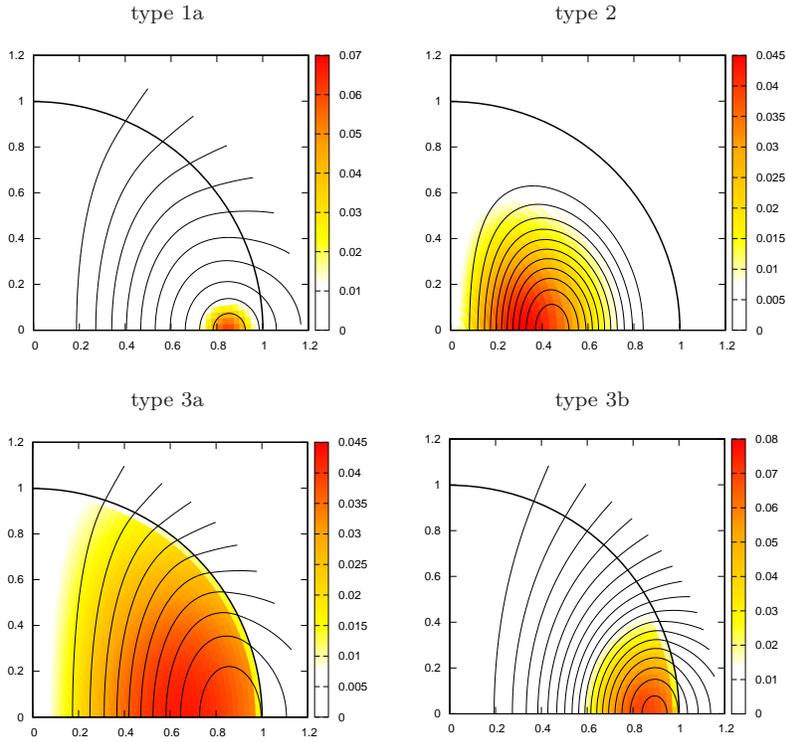}
\end{minipage}
\caption{\label{ucont_Btor}
               Comparison of various magnetic-field equilibrium
               structures. The poloidal field lines are shown in
               black, whilst the strength of the toroidal field is
               represented by the colour code. The black arc from
               $x=1$ to $y=1$ is the stellar surface. Type 1a is a
               typical `twisted-torus' configuration, where the
               toroidal field component sits inside the closed field
               lines. The other field configurations do not have this
               twisted-torus geometry. Type 2 is a confined field,
               so all poloidal field lines close within the star, whilst types
               3a and 3b are equilibria with surface currents.}
\end{center}
\end{figure}

We conclude this section by comparing the various magnetic equilibrium
models we have constructed. In figure \ref{ucont_Btor} we plot
poloidal-field lines and represent the toroidal component magnitude
with coloured shading. Since the three type-1 equilibria are rather
similar, we include only one of these (type 1a) as a representative example.

\section{Perturbations of a mixed-field star}

\subsection{Perturbation equations and numerics}

Given the apparently intractable nature of the analytic problem
\citep{taylermix}, we will study the stability of mixed
poloidal-toroidal field stars using a numerical approach. Our code
evolves the linearised MHD perturbation equations, taking the
equilibria of the previous section as background configurations. We
have successfully used this code to study instabilities in stars with
purely toroidal fields \citep{tor_instab} and purely poloidal fields
\citep{pol_instab}, and so expect its results for mixed fields to be
reliable. Since full details of our numerical scheme are reported in
\citet{tor_mode}, we content ourselves with a brief summary of the
salient details here.

We wish to study the behaviour of linear perturbations about MHD
equilibrium; this allows us to test the stability of the equilibrium,
based on whether or not we find unstable modes of the system. We work
in the Cowling approximation, where perturbations in the gravitational
potential are not evolved. This approximation is known to overestimate
the stability of a system \citep{moss_tayler}, placing some potential
doubt over any magnetic fields we find to be stable, but \emph{not} over
those systems we find to be unstable.

Instead of working directly with the velocity $\bv$ and
the perturbed magnetic field $\delta\bB$, we define flux variables
$\boldf\equiv\rho_0\bv$ and $\bbeta\equiv\rho_0\delta\bB$
(zero-subscripts denote background quantities). This allows for a
simpler, more numerically stable treatment of the stellar surface. Our
final perturbation variable is the perturbed density
$\delta\rho$. Now, in the frame corotating with the
background star, the linear perturbations of the system are described
by the following time-evolution equations\footnote{Note that in our
  earlier papers using this time evolution code, algebra errors
  affected the second and third terms of the perturbed Euler
  equation as printed. The equations used in the code were, however, correct.}:

\beqy \label{euler_magmode}
\pd{\boldf}{t} 
     &=& -\frac{\gamma P_0}{\rho_0}\nabla\drh
             + \frac{\nabla P_0}{\rho_0}\drh
             - 2\bOm\times\boldf \nn\\
     &  & - \frac{\drh}{4\pi \rho_0}(\curl\bB_0)\times\bB_0
             + \frac{1}{4\pi \rho_0} (\curl\bB_0)\times\bbeta 
             + \frac{1}{4\pi \rho_0} (\curl\bbeta)\times\bB_0
             - \frac{1}{4\pi\rho^2_0}(\nabla\rho_0\times\bbeta)\times\bB_0,
\eeqy
\beq \label{conti_magmode}
\pd{\drh}{t}=-\div\boldf,
\eeq
\beq \label{induc_magmode}
\pd{\bbeta}{t} = \curl(\boldf\times\bB_0)
                 -\frac{\nabla\rho_0}{\rho_0}\times(\boldf\times\bB_0).
\eeq

This 3D system of equations may be reduced to 2D, using the
axisymmetry of the background to decompose in the azimuthal angle
$\phi$. The desired azimuthal index $m$ is chosen at the outset
  of each evolution, which is convenient for studying the stability of
  particular modes. Imposing
the perturbations' symmetries at the poles then reduces our numerical
grid to one half-disc.  For a purely poloidal or purely toroidal
field, equatorial symmetries allow a further reduction of the grid to
a single quadrant of a disc, but these symmetries are broken \skl{in the
presence of a mixed poloidal-toroidal field, for which a half-disc
grid is required}; see the next subsection.

\skl{Each evolution is started with either an $f$- or $r$-mode
eigenfunction as initial data; each excites a different symmetry class
of perturbations (next subsection).}
The code then evolves the perturbation equations using a
predictor-corrector algorithm, and is second-order convergent. We include artificial
viscosity --- a fourth-order Kreiss-Oliger dissipation term added to
the Euler equation --- but 
take care to do so at the minimum value required to damp numerical
instabilities, so physical instabilities are minimally affected. No
corresponding artificial resistivity term was added to the induction
equation. To prevent growth of $\div\bB$ errors, we employ an
auxiliary variable and equation for divergence-cleaning
\citep{dedner}. \skl{As for the background models, we
  non-dimensionalise variables here using a combination of $G$,
  $\rho_{max}$ and $r_{eq}$. Generally we present results in
  dimensionless variables, since the important features of our
  evolutions are qualitative --- the presence or absence of dynamical
  instabilities --- rather than quantitative.}

\subsection{The importance of equatorial symmetry}

The equatorial symmetry of perturbations is closely related to the
behaviour of modes and instabilities in magnetic stars. Here we
explore this important aspect of the perturbation problem and
introduce some useful notation for later. 

Let us start with the behaviour of perturbations of an unmagnetised
fluid star. For this case, the four perturbation variables split into two
sets based on their shared symmetry properties; we denote these sets
$S_1\equiv\{ f_r,f_\phi,\drh \}$ and $S_2\equiv\{ f_\theta \}$. For example, an initial
perturbation in $f_r$ which is symmetric (or `even') about the equator will induce
corresponding symmetric perturbations in $f_\phi$ and $\drh$ but an
\emph{anti-symmetric} $f_\theta$ perturbation. A different initial perturbation,
leading to equatorially antisymmetric behaviour of the perturbations
in class $S_1$, would produce symmetric perturbations in the $S_2$
element.

From this, we can define two symmetry classes for modes (coherent global
responses of the fluid). The first we refer to as $\Pp$, and
corresponds to $S_1$ elements being even and $S_2$
elements odd; the other class, $\Pm$, is that of odd $S_1$ elements
and even $S_2$ elements. Note that the $\Pp$ class is equivalent to
the set of modes variously called polar, polar-led or spheroidal ---
for example, the $f$-mode. The $\Pm$ class is the set of modes known as
axial/axial-led/toroidal, the most familiar example being the
$r$-mode.

If we now add a purely poloidal magnetic field to the background, an
analysis of the equations shows that the original $S_1,S_2$ classes
are augmented by magnetic variables in the following manner: 
$S_1\equiv\{ f_r,f_\phi,\drh,\beta_\theta \}$ and
$S_2\equiv\{ f_\theta,\beta_r,\beta_\phi \}$. The chief instability
for a purely poloidal field is a kink mode in the closed field
line region, as described in the next subsection. This corresponds to motion in the
$\theta$ direction across the equator and hence to a \emph{symmetric}
perturbation there; an antisymmetric $v_\theta$ is zero along the
equator. For this reason the instability occurs only in the $\Pm$
symmetry class, and its development can be prevented in evolutions
which enforce $\Pp$ conditions at the equator.

In the case of a star with a purely \emph{toroidal} magnetic field,
the division of perturbations based on equatorial symmetry changes
from the poloidal-field case and the two sets become
$S_1\equiv\{ f_r,f_\phi,\drh,\beta_r,\beta_\phi \}$ and
$S_2\equiv\{ f_\theta,\beta_\theta \}$. Since $S_1$ and $S_2$ take
different forms in the purely poloidal and purely toroidal cases, it
follows that when the background field is mixed
poloidal-toroidal, the equatorial symmetry of the
perturbations will be lost.  As a result, we will no longer have a
division into $\Pp$ and $\Pm$ classes, and instead the two will be `coupled'
through the magnetic field.  Equally, there will no longer be
distinct axial-led or polar-led modes in the conventional use of the
terms.

\subsection{Diagnosing instability}
\label{instabs_theory}

\begin{figure}
\begin{center}
\begin{minipage}[c]{\linewidth}
\psfrag{r}{$\varpi$}
\psfrag{z}{$z$}
\psfrag{pc}{$\phi_c$}
\psfrag{unperturbed}{unperturbed}
\psfrag{sausage}{sausage}
\psfrag{kink}{kink}
\psfrag{mc=0}{($m_c=0$)}
\psfrag{mc=1}{($m_c=1$)}
\psfrag{location of toroidal-field}{location of toroidal-field}
\psfrag{instability (m=0,1)}{instability $(m=0,1)$}
\psfrag{location of poloidal-field}{location of poloidal-field}
\psfrag{instability (m>0)}{instability $(m>0)$}
\includegraphics[width=\linewidth]{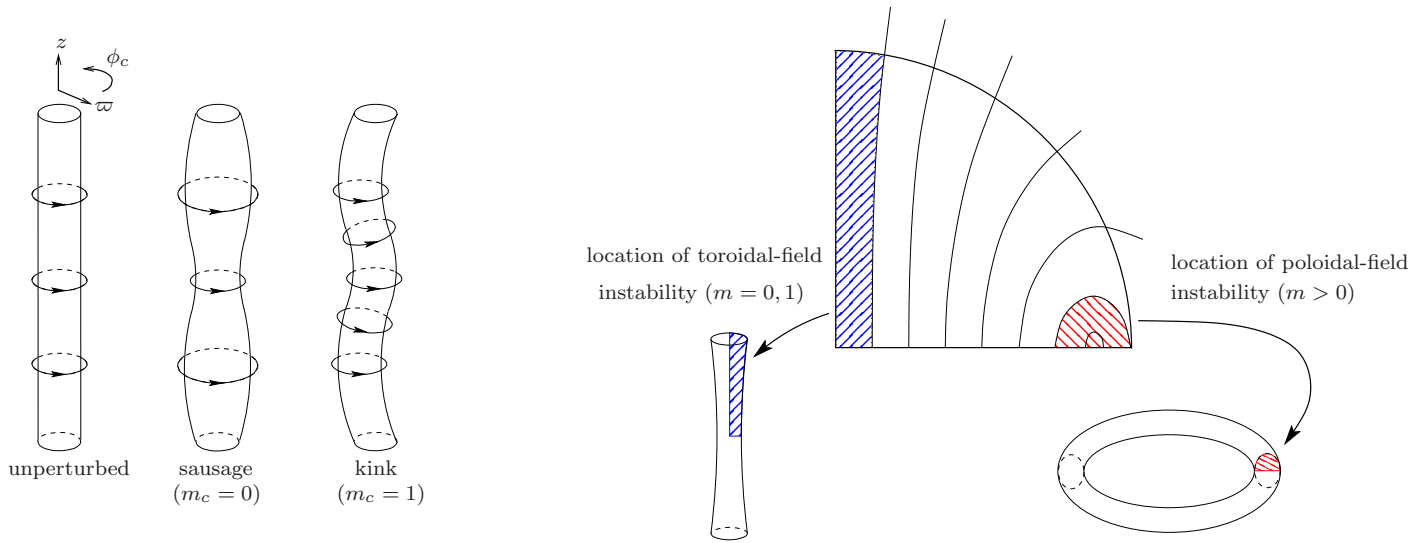}
\end{minipage}
\caption{\label{instab_locations}
               Left: instabilities of a cylindrical plasma column,
               with field lines in the $\phi_c$-direction labelled
               with arrows. The two main instabilities, in the absence
               of gravity, are the sausage/varicose mode (with
               \emph{cylindrical} azimuthal index $m_c=0$) and the kink mode
               ($m_c=1$). Right: the corresponding potentially unstable regions of a
               mixed-field star --- the cylindrical region around the pole where
               the toroidal component vanishes, and the torus around
               the neutral line, where the poloidal component
               vanishes. The potentially unstable values of $m$,
               now accounting for gravity and working in the
               spherical polar coordinates of the star, are shown.}
\end{center}
\end{figure}

Both purely poloidal and purely toroidal magnetic fields are
susceptible to rapidly-developing instabilities, of 
essentially the same origin: the unstable nature of a cylindrical
plasma column. Temporarily neglecting gravity, the simplest
instabilities of the cylindrical system are the sausage or varicose
mode and the kink mode, illustrated in figure \ref{instab_locations} (left-hand
side).  Based on this, one can identify regions of a star where
magnetic instabilities may occur: the cylindrical region around the
pole for a toroidal field, and the torus around a poloidal field's
neutral line (i.e. the line along which the poloidal field
vanishes). \skl{Any such instabilities will be related to the geometry of the
magnetic field, so should be present even for weak fields, albeit with slower
growth rates \citep{taylerpol,wright,taylertor}.}

We now turn to the right-hand side of figure \ref{instab_locations},
and consider first a star with a purely toroidal magnetic field. The geometry is
similar to that of the cylindrical plasma, but the star's self-gravity
must now be accounted for. Since both sausage and kink instabilities in this
case can operate parallel to shells of constant gravitational
potential, however, neither should be suppressed; accordingly, we expect
$m=0$ and $m=1$ magnetic instabilities to exist\footnote{In the
  special case of an incompressible star, the sausage mode \emph{does}
  have to move fluid elements along the $z$-axis and hence will be
  suppressed by gravity, but the $m=1$ instability will remain.}.
In a poloidal field the 
geometry is different, with the unstable cylinder closed into a torus,
resulting in potential instability in all $m>0$ perturbations. The
sausage mode of the cylinder, on the other hand, is suppressed by
gravity; so we do not expect a significant $m=0$ instability for a
poloidal-field star.

For a mixed poloidal-toroidal field to be stable, we need to check
that all instabilities that occur for a purely poloidal or toroidal
field are eliminated in the mixed configuration \skl{(and that no new
instabilities arise)}.
In this paper, we only consider non-axisymmetric modes. To test for any
instabilities due to the toroidal component, therefore, we look at the behaviour
of $m=1$ perturbations in the vicinity of the pole. By contrast, we anticipate that
instabilities of the poloidal component will be localised in the
closed-field line region, for any $m\geq 1$. For a purely poloidal field only $\Pm$
initial data will produce these (kink) instabilities, but a
mixed field allows for $\Pp$--$\Pm$ coupling --- so that any initial
data may result in the excitation of a poloidal-component
instability.

In the initial phase of an instability, an unstable mode is excited
with an amplitude which grows exponentially in time; to study this, it is
sufficient to consider the behaviour of linear perturbations about an
equilibrium.  Beyond some amplitude the mode `saturates', causing a
nonlinear rearrangement of the magnetic field for which the perturbative
regime is no longer applicable. Since we only evolve the linearised
system, however, the background is stationary and experiences no
reduction in magnetic energy as the amplitude of a perturbation grows
--- hence the instability is able to grow indefinitely. This is clearly not
realistic, but it does allow us to determine the growth rate of the
initial instability with ease.
In all cases, the onset time of the instability is expected to
correspond to the time taken for an Alfv\'en wave to cross the
relevant part of the system. As a diagnostic, we define the following
quantity as an order of magnitude estimate for this time:
\beq \label{alf_time}
\tau_A\equiv\frac{R_*}{\bar{c}_A}=R_*\sqrt{\frac{4\pi\bar\rho}{\bar{B}^2}},
\eeq
where $c_A$ is the Alfv\'en speed and overbars denote
volume-averaged quantities. To ensure we compare configurations of the
same physical field strength, we redimensionalise to typical neutron star
parameters (a mass of $1.4M_\odot$ and a radius $R_*=10$ km) and take
a field strength of $\bar B=3\times 10^{16}$ G as our canonical value.
The corresponding (dimensionless) Alfv\'en timescale
$\hat{\tau}_A\approx 40$. This strong magnetic field and 
correspondingly short dynamical timescale allows for faster numerical
evolutions, but we have confirmed our results also hold at lower field
strengths. In addition, we have checked that the growth rate of any unstable
mode converges with numerical grid resolution, and scales approximately linearly
with the magnetic field strength.

Note that confirming the stability of a magnetic field is far more
difficult than showing it is unstable: for the latter one need only
find a particular mode with exponential growth in time, whilst for the
former one must establish that \emph{every} potential perturbation
results only in innocuous oscillations about the equilibrium and not
unstable growth. Within the limitations of our numerics, we are able
to test the stability of equilibria against nonaxisymmetric initial
perturbations with $\Pp$ or $\Pm$ symmetry, up to
$m=6$. Since this test easily shows up instabilities of purely
poloidal and purely toroidal fields \citep{tor_instab,pol_instab}, we
may have some confidence that a mixed poloidal-toroidal field showing
no such instability is `stable'. More accurately, it will not be
susceptible to the fastest-growing class of instabilities and is
likely to be dynamically stable; we cannot establish its stability on
secular timescales.

\section{Stability analysis of mixed-field stars}

\subsection{Type 1 (twisted-torus) equilibria}

\begin{figure}
\begin{center}
\begin{minipage}[c]{\linewidth}
\psfrag{deltaM}{\large{$\delta\hat{\clE}_{\textrm{mag}}$}}
\psfrag{t}{\large{$\hat{t}$}}
\psfrag{m=1 P+}{\small{$m=1,\ \Pp$}}
\psfrag{m=1 P-}{\small{$m=1,\ \Pm$}}
\psfrag{m=2 P+}{\small{$m=2,\ \Pp$}}
\psfrag{m=2 P-}{\small{$m=2,\ \Pm$}}
\psfrag{poloidal-field background}{poloidal-field background}
\psfrag{mixed-field background}{mixed-field background}
\includegraphics[width=\linewidth]{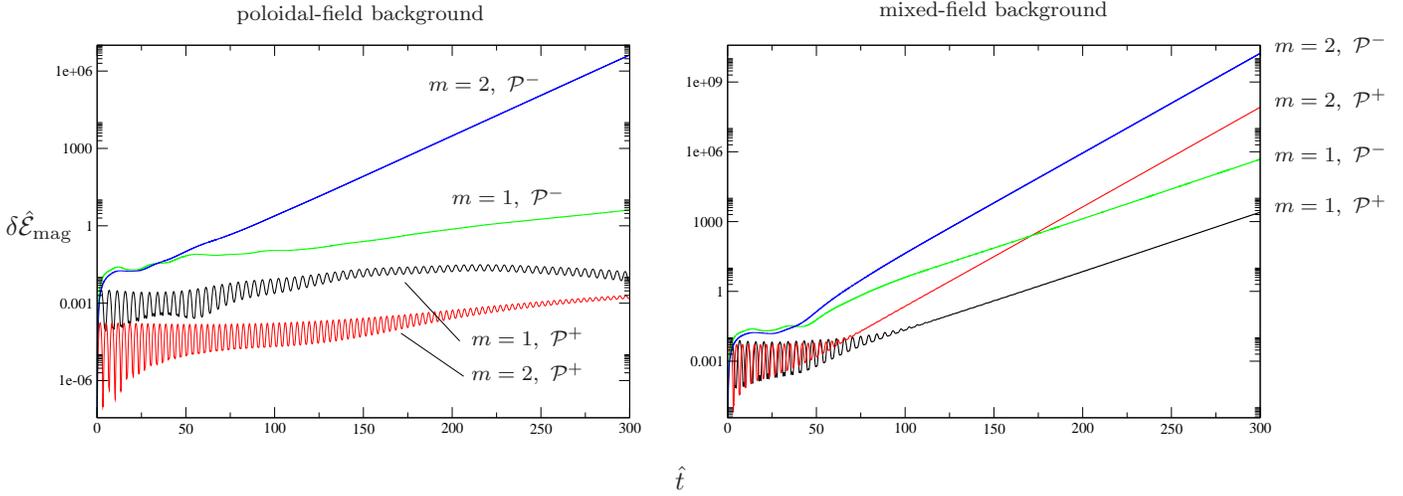}
\end{minipage}
\caption{\label{type1_instab}
              The instability of type-1 (twisted-torus) equilibria; results are shown
              for type 1a, but the corresponding plots for types 1b
              and 1c are very similar. \skl{Plotting the evolution of
              the perturbed magnetic energy,} we compare the behaviour of
              $m=1$ and $2$ initial perturbations (with $\Pp$ and
              $\Pm$ symmetry) in a star with a purely
              poloidal field (left), and in one where a toroidal
              component has been added (right). \skl{In both cases,
                unstable growth is seen after approximately one
                Alfv\'en timescale ($\hat{t}\approx 40$ in these
                dimensionless variables).} For the mixed-field
              star $\etor/\emag=3\%$; this is a high percentage for this
              class of equilibrium, but nonetheless does not produce
              any stabilisation. For a mixed field, 
              magnetic perturbations across the equator have no
              definite equatorial symmetry; in a sense the mixed field allows for
              `mixing' of the two $\mathcal{P}$-classes. For this
              reason, although $\Pp$ initial data starts by evolving
              stably, as for a poloidal-field background, at later
              times energy is transferred to the unstable class of
              perturbations.}
\end{center}
\end{figure}

We now begin our analysis of the magnetic stability of barotropic
stars containing both poloidal and toroidal field components, using the
various background models discussed in section 2. In particular, we
are testing for the possible presence of any instabilities known from
the purely-poloidal and purely-toroidal field cases. In this
subsection we test the stability of twisted-torus equilibria (type 1),
as constructed in section \ref{tt_eqa}. These equilibria are
poloidal-dominated, with a toroidal component only occupying the small
volume within the closed-field line region.

In figure \ref{type1_instab} we compare the stability of a type-1a
mixed field with that of a poloidal field. It would be reasonable to
expect the addition of a toroidal component to remove, or at least
reduce, the instability of the poloidal field \citep{wright}; despite this,
our results show \emph{no such} stabilisation. On the left-hand side
we plot the evolution of perturbed magnetic energy for $\Pp$
(polar) and $\Pm$ (axial) perturbations, on a poloidal-field
background. As expected from section \ref{instabs_theory}, the $\Pp$ class evolve
stably, whilst the $\Pm$ class suffer an instability that appears at
$\hat{t}\approx 30-50$; cf. our Alfv\'en timescale estimate of
$\hat\tau_A\approx 40$ for this field strength ($\bar B=3\times 10^{16}$ G).

The right-hand plot shows the corresponding evolutions for a
mixed-field star with $\etor/\emag=3\%$, relatively high for a type-1
equilibrium. As for the poloidal field, after one Alfv\'en timescale
the perturbed magnetic energy grows exponentially; furthermore, since the
mixed field breaks the equatorial symmetry of the perturbations (see
section \ref{instabs_theory}), even $\Pp$ initial data results in an
excitation of unstable modes. The growth rate of the $m=2$ instability
is not reduced by the addition of a toroidal component, nor are the
$m=4$ or $6$ instabilities (omitted in figure \ref{type1_instab} to
avoid cluttering it). The $m=1$ instability, in fact, grows
\emph{faster} for a mixed field than a purely poloidal one. Finally,
although we have only plotted results for type-1a equilibria, type-1b
and 1c mixed fields are similarly unstable.

\begin{figure}
\begin{center}
\begin{minipage}[c]{0.5\linewidth}
\psfrag{poloidal}{poloidal}
\psfrag{mixed}{mixed}
\includegraphics[width=\linewidth]{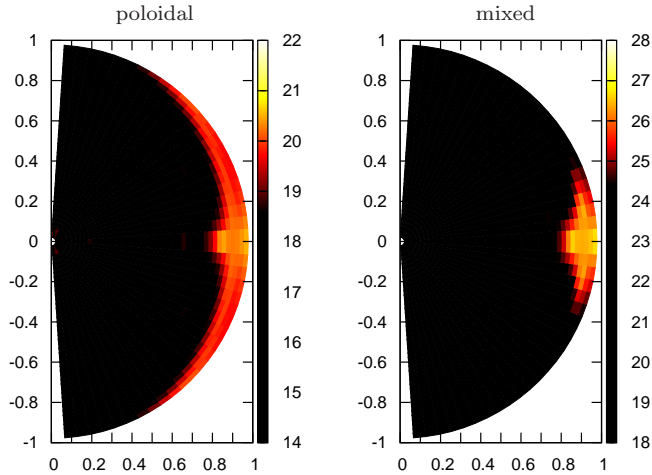}
\end{minipage}
\caption{\label{Bpert}
               The magnitude of the perturbed magnetic field $|\delta\bB|$,
               in logarithmic scale, at late times in the evolution of
               two stars. The left-hand plot is for $\Pm$
               perturbations on a poloidal-field background, whilst the
               right-hand plot shows perturbations of a mixed field
               (for which there is no equatorial symmetry); $m=2$ for both
               cases. The poloidal-field instability --- manifested as
               exponential growth around the neutral line --- is seen to operate
               in the mixed-field star too. The corresponding plot for
               $m=1$ shows similar growth around the neutral line too,
               \skl{with no evidence for 
               additional instabilities originating from the toroidal
               component of the mixed field.}}
\end{center}
\end{figure}

\begin{figure}
\begin{center}
\begin{minipage}[c]{0.9\linewidth}
\psfrag{vth-vr}{$\log\left|\displaystyle{\frac{v_\theta}{v_r}}\right|$}
\psfrag{vth-vphi}{$\log\left|\displaystyle{\frac{v_\theta}{v_\phi}}\right|$}
\psfrag{t}{$\hat{t}$}
\psfrag{poloidal}{poloidal}
\psfrag{mixed}{mixed}
\includegraphics[width=\linewidth]{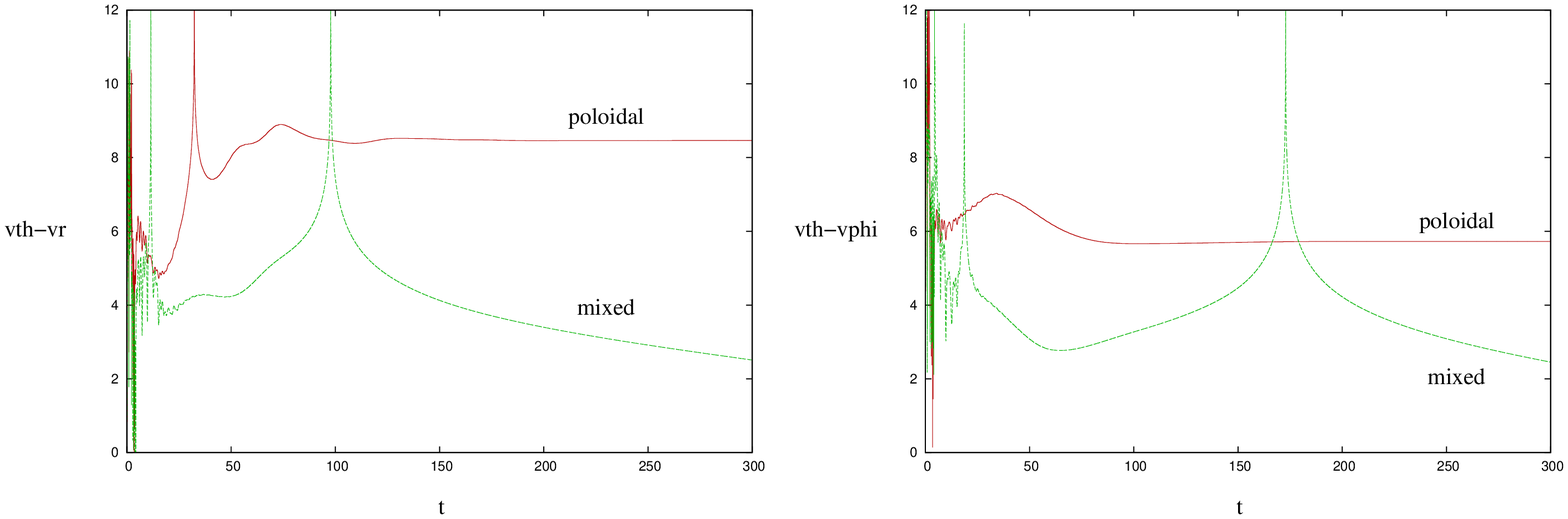}
\end{minipage}
\caption{\label{vratio}
               \skl{Confirmation that the instability of poloidal and type-1
                 mixed-field equilibria involves predominantly
                 $\theta$-direction motion. We compare the magnitude
                 of $v_\theta$ with that of the other fluid velocity
                 components, for a point in the closed-field line
                 region. The $r$ and $\phi$ components of the velocity
                 are more significant in the mixed-field configuration
                 than for a purely poloidal-field background, but are
                 still two orders of magnitude weaker than the
                 $\theta$-component.}}
\end{center}
\end{figure}

\skl{Two additional tests help us to confirm that the instability from
  figure \ref{type1_instab} is due to the poloidal component.}
First, we look at its location 
within the star. In figure \ref{Bpert} we plot the magnitude of the
perturbed field $|\delta\bB|$, in the case of a poloidal field and a
type-1 mixed configuration. The plots are in logarithmic scale and
show the late stages of the evolution, where the unstable mode is
completely dominant. For both field configurations shown, there is
clear exponential growth around the neutral line --- slightly more
localised for the mixed field. Although we only show results for
$m=2$, the unstable growth occurs in the same location for $m=1$,
\skl{with no indication in this latter case of additional growth due
  to the potentially unstable toroidal component}.

\skl{The poloidal-field instability discussed in section \ref{instabs_theory}
  should manifest itself as $\theta$-direction motion localised in the
  closed-field line region. Having confirmed that our instability
  appears in the expected location, our final test is to study the fluid
  motion in this region. In figure \ref{vratio} we take a point in the
  closed-field line region and plot the local ratios of $v_\theta$ to $v_r$
  and $v_\theta$ to $v_\phi$ over time. The behaviour is consistent with our
  expectations: after the onset of instability, the
  $\theta$-direction motion is always more than two orders of magnitude
  greater than that in either of the other directions.}

\subsection{Type 2 (confined-field) equilibria}

\begin{figure}
\begin{center}
\begin{minipage}[c]{0.8\linewidth}
\psfrag{deltaM}{$\delta\hat{\clE}_{\textrm{mag}}$}
\psfrag{t}{$\hat{t}$}
\psfrag{m=1}{\small{$m=1$}}
\psfrag{m=2}{\small{$m=2$}}
\psfrag{m=4}{\small{$m=4$}}
\psfrag{m=6}{\small{$m=6$}}
\includegraphics[width=\linewidth]{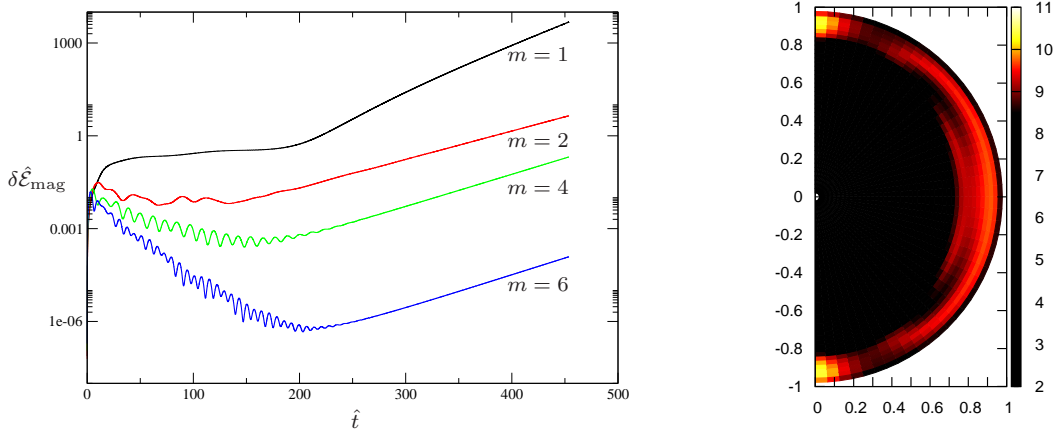}
\end{minipage}
\caption{\label{type2_instab}
              Left: the instability of a confined-field (type 2)
              equilibrium, seen from the exponential growth of the
              perturbed magnetic energy in time. The background stellar model is very
              different from type 1: it is toroidal-dominated
              ($\etor/\emag=65\%)$ and does not have a twisted-torus
              geometry. Although the field strength is the same as
              before ($\bar B=3\times 10^{16}$ G), no instabilities
              manifest themselves until $\hat{t}\approx 200$ --- five
              times our estimated Alfv\'en timescale. Right: the
              unstable growth in $|\delta\bB|$ is fastest around the
              pole, suggesting that the toroidal component is the main
              origin of the instability for type-2 equilibria.}
\end{center}
\end{figure}

Next we turn to a stability analysis for a star with a confined
magnetic field, as described in section \ref{cf_eqa}. This model
does not have a twisted-torus structure and is not poloidal-dominated;
65\% of its magnetic energy comes from the toroidal component. These
differences allow us to explore whether the twisted-torus structure or
dominant poloidal field are responsible for the instability of type-1
equilibria.

On the left-hand side of figure \ref{type2_instab} we plot the evolution of perturbed
magnetic energy for $m=1,2,4$ and $6$ azimuthal indices. Although the
$m=2,4$ and $6$ instabilities still appear to exist for this magnetic
field, their growth rates are greatly reduced in comparison with the
type-1 equilibria results. In addition, the onset time for the
instabilities is rather longer: about five times that for type 1. One
similarity between the type-1 and type-2 results is that the $m=1$
instability seems to grow more quickly for a mixed field than a purely
poloidal one.

Since the type-2 equilibrium is toroidal-dominated, we wish to see
which field component is responsible for its instability. Accordingly,
on the right-hand side of figure \ref{type2_instab}
we plot $|\delta\bB|$ throughout the star for an $m=1$ evolution. We
see the fastest unstable growth occurring near the polar surface,
indicating that the toroidal component is now unstable (see section
\ref{instabs_theory}). There is also 
substantial growth away from the pole, however, and $m>1$
perturbations are seen to be unstable too; both of these results
indicate that the poloidal component also plays a role in the instability
of this field type.

\subsection{Type 3 (surface-current) equilibria}

In this subsection we test the stability of type 3 equilibria, our
final class of mixed-field configurations,
where we have an exterior poloidal field but nonetheless a strong
toroidal component. This is achieved by allowing for a step in the
toroidal component at the surface and requires a corresponding surface
current. Recall, however, that we are not able to study the surface
dynamics of the perturbed magnetic field directly, since we evolve the flux
variable $\bbeta=\rho_0\delta\bB$.

\begin{figure}
\begin{center}
\begin{minipage}[c]{\linewidth}
\psfrag{deltaM}{\large{$\delta\hat{\clE}_{\textrm{mag}}$}}
\psfrag{t}{\large{$\hat{t}$}}
\psfrag{type 3a}{type 3a}
\psfrag{type 3b}{type 3b}
\psfrag{m=1}{\small{$m=1$}}
\psfrag{m=2}{\small{$m=2$}}
\psfrag{m=4}{\small{$m=4$}}
\psfrag{m=6}{\small{$m=6$}}
\includegraphics[width=\linewidth]{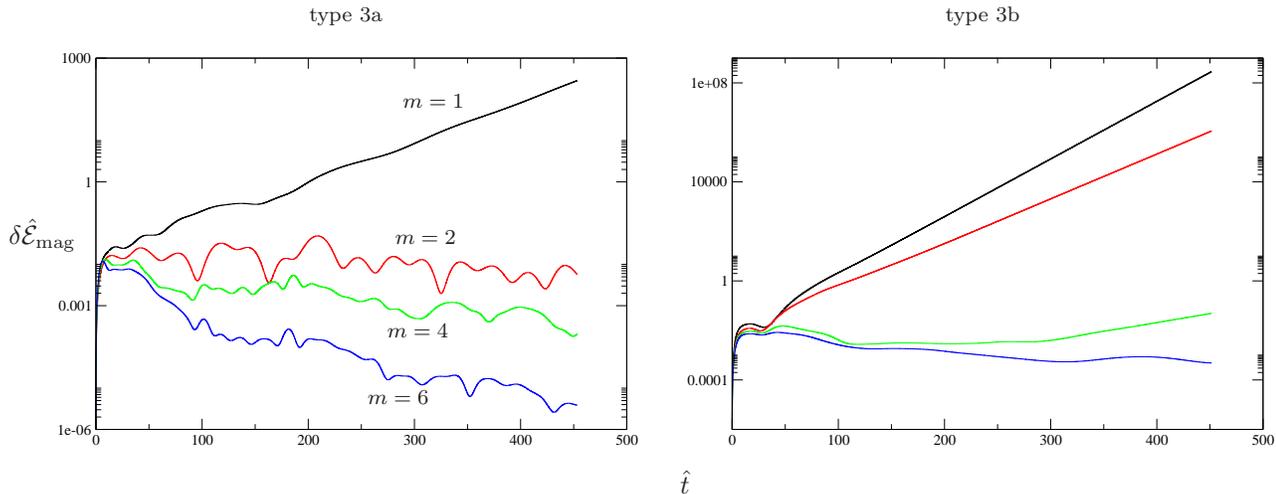}
\end{minipage}
\caption{\label{type3_instab}
               Behaviour of $m=1,2,4,6$ perturbations on background
               equilibria with surface currents and fairly high
               percentages of toroidal-field energy ($50\%$ for type 3a, $30\%$ for type 3b).
               We see that higher-$m$ modes are stabilised but the 
               $m=1$ poloidal-field instability is not removed even in the case of
               50\% toroidal energy.}
\end{center}
\end{figure}

We consider two examples of surface-current equilibria, type 3a and
type 3b, as described in section \ref{sc_eqa}. Both have substantial
toroidal components, with comparable energy to the poloidal component:
for type 3a $\etor/\emag=50\%$, and for type 3b
$\etor/\emag=30\%$. In each case, the value was the upper limit on the
percentage of toroidal field possible with our chosen functional forms.
From figure \ref{type3_instab} we see some degree of stabilisation in
these equilibria, compared with those of type 1. In the evolution for
type 3b, the $m=6$ instability appears to have been removed and the
$m=4$ instability considerably reduced. However, the $m=1$ and $2$
instabilities remain.

Type 3a is the closest to a stable barotropic equilibrium we have
found, with no evidence for the existence of $m=2,4$ or $6$
instabilities. Despite this, we still see exponential growth for $m=1$
--- which is enough to render this magnetic field unstable. Given the
stabilisation of other modes, one might question whether the $m=1$
growth is numerical rather than physical in origin. We have checked
this, as for all other unstable configurations, by convergence-testing
the instability growth rate. Furthermore, the unstable growth was localised
around the neutral line, closely resembling the plots in figure
\ref{Bpert}. This suggests an instability of physical origin, related
to the poloidal component of the magnetic field, even though the
toroidal component contributes $50\%$ of the magnetic energy.

\subsection{Rotation}

We conclude our stability analysis by looking at the effect of
rotation. Rotation may act to reduce magnetic instabilities, though
there is some uncertainty in the literature about how effective the
stabilisation will be. Part of the problem is that there are no
conclusive analytic results for the rotating case, due to the
increased complexity of the stability conditions
\citep{frieman,lynd_ostr,rotmag_stab} with respect to the static case
\citep{bernstein}.

Although no definitive analytic results exist for the stability of rotating
magnetised stars, the work of \citet{pitts_tayler} provides some
suggestions, based on approximate treatments of some model
problems. They predict that rapid rotation may remove magnetic instabilities
already present in the non-rotating star, but that a new instability
will be introduced, albeit one with a lower growth rate. From
numerical studies, there seems to be agreement that rotation can
reduce toroidal-field instabilities
\citep{braith_tor,kitch,tor_instab,kiyoshi}, though not necessarily
provide complete stabilisation. For poloidal fields,
\citet{gepp_rhein} and \citet{pol_instab} find that rotation has some
stabilising effect, whilst \citet{braith_pol} does not. With somewhat
tentative expectations, we now turn to our results for a rotating star
with a mixed poloidal-toroidal field. We specialise to $m=2$
perturbations on a type-1a mixed-field background in this subsection,
but the results are representative of other field types and
perturbations we have tested.

\begin{figure}
\begin{center}
\begin{minipage}[c]{\linewidth}
\psfrag{emag}{$\delta\hat{\clE}_{\textrm{mag}}$}
\psfrag{t}{$\hat{t}$}
\psfrag{B3}{$\bar B=3\times 10^{16}$ G}
\psfrag{B225}{$\bar B=2.25\times 10^{16}$ G}
\psfrag{om0}{$\hat\Omega=0$}
\psfrag{om0891}{$\hat\Omega=0.089$}
\psfrag{om1312}{$\hat\Omega=0.131$}
\psfrag{om4117}{$\hat\Omega=0.412$}
\psfrag{omega}{$\hat\Omega$}
\psfrag{zeta}{$\hat\zeta$}
\includegraphics[width=\linewidth]{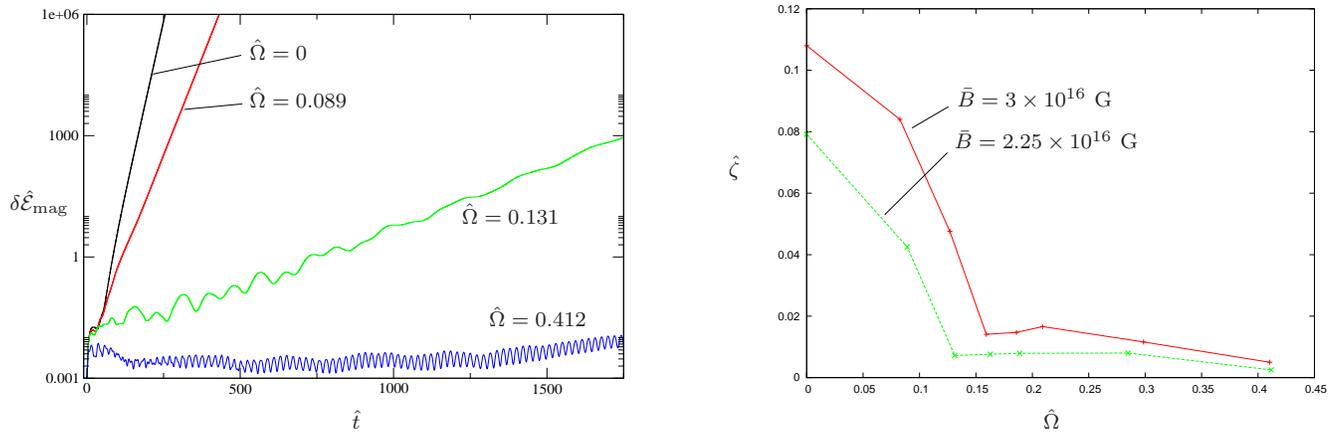}
\end{minipage}
\caption{\label{rot_plot}
               Left: the evolution of the perturbed magnetic energy in
               a rotating mixed-field star, for different
               dimensionless rotation rates. Rotation is seen to
               reduce, but not remove, the instability. Note that the
               Keplerian frequency in these units $\hat\Omega_K\approx
               0.72$. Numerical instability prevents us from studying
               more rapidly-rotating stellar models than
               $\hat\Omega=0.412$, around 57\% of the Keplerian
               value. Right: instability growth rate $\zeta$ as a
               function of rotation rate, for two field
               strengths. The instability present for $\Omega=0$ seems
               to be reduced by rotation, but at around
               $\hat\Omega=0.15$ the lines plateau, suggesting the
               appearance of a second instability.}
\end{center}
\end{figure}

To study the effect of rotation on magnetic instabilities, we monitor
the evolution of the perturbed magnetic energy once again.  We
quantify the (exponential) instability growth rate using the
parameter
\beq
\zeta\equiv\frac{1}{\Delta t}\Delta
                        \left[ \ln\brac{\frac{\delta\emag}{{\emag}_0}}
                        \right].
\eeq
From the
left-hand plot of figure \ref{rot_plot} we find that rotation
certainly decreases the growth rate of the perturbations, although
some degree of unstable growth is present in all results. Our code suffers some
numerical instabilities for very rapidly-rotating stars, so the
highest rotation rate we include here is $57\%$ of the Keplerian
value. 

The work of \citet{pitts_tayler} suggests that a non-rotating star's magnetic
instabilities should be suppressed when the rotational
angular velocity exceeds the corresponding Alfv\'en velocity
$\Omega_A\approx 2\pi/\tau_A$. Using our previous estimate for
$\tau_A$ \eqref{alf_time}, we expect this suppression to occur at $\hat\Omega\approx
0.11$ for $\bar B=2.25\times 10^{16}$ G and at $\hat\Omega\approx
0.14$ for $\bar B=3\times 10^{16}$ G.  As the original instability is
removed, however, \citet{pitts_tayler} predict a new instability will
be induced whose growth rate may be slower but still significant.

With these results in mind, we now turn to the right-hand plot of
figure \ref{rot_plot}. Here we quantify the effect 
of rotation, plotting the rate of exponential growth of the magnetic
energy as a function of rotation rate. We consider stars with
$\bar B=2.25$ and $3\times 10^{16}$ G (a ratio of 3:4), to
investigate the scaling of instability growth rate with
field strength. At zero rotation rate the more highly magnetised star
has a growth rate $4/3$ that of the other, confirming the linear
dependence we expect in this case. Adding rotation causes an initial
decrease in growth rate which is roughly linear in $\Omega$, up to
some threshold value: $\hat\Omega\approx 0.13$ for the weaker magnetic
field, $\hat\Omega\approx 0.16$ for the stronger field. These values
agree rather well with the estimates of the previous paragraph. Beyond
the threshold rotation rate for each star, the unstable growth rate seems to reach a
plateau value of $\sim 10\%$ that of the non-rotating case. We believe
this corresponds to the new magneto-inertial instability predicted to
occur for sufficiently rapid rotation.

\vspace{-0.4cm}
\section{Discussion: possible stable equilibria}

The aim of this paper has been to construct as wide a range of mixed
poloidal-toroidal magnetic fields as possible in barotropic stars, and
then to test their stability. We have found that \emph{all} of these
equilibria are unstable: poloidal-dominated twisted-torus models
(like those of \citet{tomi_eri,LJ09,ciolfi_eqm}), toroidal-dominated fields confined
within the star \citep{haskell,duez} and equilibria with surface
currents \citep{colaiuda}. For equilibria where $\etor/\emag\leq 50\%$
the instability's origin appeared to be the poloidal component; for the
confined-field model we tested, with $\etor/\emag=65\%$, we found
evidence of a toroidal-field instability, but some indications of
instability in the poloidal component too. This seems to leave little
scope for configurations where all magnetic instabilities are
suppressed. On the other hand, we have observational 
evidence for long-lived stellar magnetic fields, so stable magnetic
equilibria clearly exist. Here we discuss possible mechanisms to
suppress the instabilities we have found.

{\bf Rotation:} highly magnetised stars (like many Ap/Bp stars,
magnetic white dwarfs and magnetars) often rotate relatively slowly. In some,
however, the kinetic rotational energy greatly 
exceeds the magnetic energy and one might expect suppression of
magnetic instabilities. We find, however, that when rotation seems rapid
enough to suppress the original magnetic instability (of a
non-rotating star), a new instability appears to be introduced. This new
magneto-inertial instability has a growth rate of $\sim 10\%$ the
non-rotating value --- still very short on stellar timescales.

{\bf Non-axisymmetry:} our code uses a decomposition of the
perturbation equations in the azimuthal angle, which relies on having an
axisymmetric background star. It is possible that a
non-axisymmetric magnetic field will not suffer from the same
instabilities we have found. Indeed, the non-linear evolutions of \citet{lasky} and
\citet{ciolfi_nonlin} seem to show unstable initial fields developing
non-axisymmetric structure. At the end of these reported simulations,
however, the field remains highly dynamic in the closed-field line
region, so it is probably too early to regard the results as
stable equilibria. It would be very interesting to see whether these
fields settle down over longer evolution times.

{\bf Stratification:} perhaps the easiest way to explain our results
is to say that they are irrelevant for real stars, where the equation
of state is not barotropic but instead stratified. Indeed, the only
numerical evolutions which seem to produce stable magnetic equilibria
use a stratified stellar model, where the stratification is due to
temperature/entropy gradients
\citep{braith_nord,braith_torpol}. Whether these equilibria are, in
fact, stable is not completely settled: \citet{bon_urp} discuss a generic
magnetic instability whose lengthscale can be very short, and
hence not show up in numerical simulations (for which the grid is
necessarily relatively coarse).

In any case, results for entropy-based
stratification may not be applicable to neutron stars, where the
stratification comes from composition gradients.
Recently, some first attempts have been made to model such
equilibria \citep{mastrano,GAL,LAG}, but these are qualitatively similar to barotropic
results and their stability is unknown. Note that there
is no guarantee that \emph{radial} stratification will suppress the
instabilities we find, which mainly involve $\theta$-direction motion
(although it does appear to in \citet{braith_nord} and related papers).

{\bf Elastic crust:} in the case of neutron stars, the outermost
layer of the star (around a kilometre in thickness) consists of an
elastic crust. This crust \emph{does} have the potential to suppress
unstable motion of the kind studied in this paper, up to some critical
field strength roughly corresponding to when the magnetic energy
exceeds the crustal elastic energy. The precise value depends on the
poorly-known yield strain of the crust, but is estimated at around
$10^{14}$ G \citep{thom_dunc93}. Above this field strength, a 
magnetic instability can cause the crust to crack, which is a
plausible scenario for the triggering of magnetar bursts
\citep{thom_dunc95}. For white dwarfs and Ap/Bp stars, there is no
analogous outer region to hold an unstable field in place.

{\bf Superconductivity:} relatively little is known about the
equilibria and stability of magnetic fields in superconducting
stars. The magnetic force takes a very different form from the normal
Lorentz force, however, so it is not unreasonable to expect
significant differences in the magnetic field configurations
too \citep{mendell98,GAS}. Neutron stars quickly become cool enough to contain
superconducting regions \citep{page,shternin}, so this aspect of their
physics may help to stabilise (or destabilise!) their magnetic fields.

\vspace{-0.6cm}
\section{Summary}

We construct a wide range of barotropic stellar equilibria, with mixed
poloidal-toroidal magnetic fields. These include poloidal-dominated
and toroidal-dominated configurations, as well as those where both
components have comparable energies. Every field configuration is found to
suffer instabilities whose origin appears to be one of the two field
components. Starting with a poloidal-dominated field, increasing the
strength of the toroidal component suppresses unstable
poloidal-component modes of higher azimuthal number, but the $m=1$
instability is difficult to remove. For a sufficiently strong
toroidal component, the origin of the $m=1$ instability appears to change from
the poloidal to the toroidal component. Rotation
reduces the instability growth rate, but seems to introduce a new
instability of slightly weaker growth rate (around 10\% of the
non-rotating value).

Although this paper does not \emph{prove} the non-existence of
stable fields in barotropic stars, it suggests that any such fields
form --- at best --- a very restricted set (which we have been unable to obtain). By
contrast, many classes of star are observed to have long-lived
magnetic fields, so we expect that stable magnetic equilibria do
exist. Our results indicate that barotropic models may, therefore, be of
limited astrophysical relevance. For main-sequence stellar models,
the addition of entropy-gradient stratification seems to allow for
stable equilibria, and the same may then be true for white dwarfs. For the
modelling of neutron star fields, we 
believe that it is now particularly important to account for
additional physics beyond the barotropic-fluid model, including the effect of a
crust, composition gradients and superconductivity.

\vspace{-0.4cm}
\section*{Acknowledgments}

We thank Kostas Glampedakis for useful discussions during this project and
Bryn Haskell, Paul Lasky and Burkhard Zink for clarifying aspects
of their work. We are also grateful to the referee for their helpful
criticism. SKL is supported by the German Science Foundation (DFG) via SFB/TR7;
DIJ by STFC via grant number ST/H002359/1. We acknowledge additional
support from CompStar, an ESF research networking programme.

\label{lastpage}

\end{document}